\documentclass[10pt,final,journal]{IEEEtran}

\usepackage{cite}
\usepackage{amsmath,amssymb,amsfonts}
\usepackage{algorithmic}
\usepackage{graphicx}
\usepackage{textcomp}
\usepackage{xcolor}
\usepackage{multicol}
\usepackage{multirow}
\usepackage{afterpage}
\usepackage{siunitx}

\def\BibTeX{{\rm B\kern-.05em{\sc i\kern-.025em b}\kern-.08em
    T\kern-.1667em\lower.7ex\hbox{E}\kern-.125emX}}

\begin{document}
%
\title{SFOL DME Pulse Shaping Through Digital Predistortion for High-Accuracy DME}
%
%
%

        
\author{
Sunghwa~Lee, Euiho~Kim,~\IEEEmembership{Member,~IEEE}, and Jiwon~Seo,~\IEEEmembership{Member,~IEEE}
\thanks{Manuscript received June 00, 2021; }
\thanks{This work was supported by the Korea Agency for Infrastructure Technology Advancement (KAIA) grant funded by the Ministry of Land, Infrastructure and Transport (Grant 21TBIP-C155921-02).}
\thanks{Authors' addresses: S.~Lee and J.~Seo are with the School of Integrated Technology, Yonsei University, Incheon 21983, Republic of Korea, E-mail: (sunghwa.lee@yonsei.ac.kr; jiwon.seo@yonsei.ac.kr); E.~Kim is with the Department of Mechanical and System Design Engineering, Hongik University, Seoul 04055, Republic of Korea, E-mail: (euihokim@hongik.ac.kr). \textit{(Corresponding authors: Euiho Kim; Jiwon Seo.)}}
}

%
%

%
\markboth{IEEE Transactions on Aerospace and Electronic Systems,~Vol.~00, No.~0, June~2021}%
{Lee \MakeLowercase{\textit{et al.}}: SFOL DME Pulse Shaping Through Digital Predistortion for High-Accuracy DME}

\maketitle

\begin{abstract}
The Stretched-FrOnt-Leg (SFOL) pulse is a high-accuracy distance measuring equipment (DME) pulse developed to support alternative positioning and navigation for aircraft during global navigation satellite system outages. To facilitate the use of the SFOL pulse, it is best to use legacy DMEs that are already deployed to transmit the SFOL pulse, rather than the current Gaussian pulse, through software changes only. When attempting to transmit the SFOL pulse in legacy DMEs, the greatest challenge is the pulse shape distortion caused by the pulse-shaping circuits and power amplifiers in the transmission unit such that the original SFOL pulse shape is no longer preserved. This letter proposes an inverse-learning-based DME digital predistortion method and presents successfully transmitted SFOL pulses from a testbed based on a commercial legacy DME that was designed to transmit Gaussian pulses.
\end{abstract}


%
\IEEEpeerreviewmaketitle

\section{Introduction}
%
%
%
%
\IEEEPARstart{G}{lobal} navigation satellite systems (GNSS) are key infrastructure elements for modern aircraft navigation and air traffic control systems. However, a GNSS service can be easily disrupted by man-made signals, such as in radio frequency interference or jamming \cite{Park19:Single}. To secure the seamless operation of the aircraft navigation and air traffic control during GNSS signal disruptions in local or wide area, the Federal Aviation Administration (FAA) of the U.S. has been searching for possible Alternative Position, Navigation, and Timing (APNT) systems since approximately 2010. 

Various candidate APNT systems have been proposed, including wide area multilateration (WAM) using Automatic Dependent Surveillance-Broadcast (ADS-B) signals \cite{Chen13}, enhanced distance measuring equipment (DME) using carrier-phase measurement for ranging \cite{Pelgrum12}, and L-band digital aeronautical communication system type 1 (LDACS1) \cite{Osechas16}. Rather than establishing a new APNT system, the FAA decided to use DME/DME as a short-term backup solution for GNSS and has been expanding the existing DME ground networks \cite{FAA16}. DME/DME is a positioning method that computes the horizontal position of an aircraft using more than two slant ranges from DME ground transponders. The DMEs currently used in DME/DME are conventional Gaussian-pulse-based DMEs.

However, DME/DME systems based on the Gaussian pulse may have large positioning errors because the naive Gaussian pulse is subject to multipath-induced ranging errors over 100 m, which is not sufficient for an ultimate long-term APNT solution. Efforts have been made to develop alternative DME pulse waveforms that overcome the shortcomings of the Gaussian pulse while simultaneously meeting the current DME pulse shape specifications from the International Civil Aviation Organization (ICAO). Among the various alternative DME pulses proposed \cite{Kim17a, Kim17b}, the Stretched-FrOnt-Leg (SFOL) pulse designed by applying genetic algorithms is known to have the best performance.

To improve DME/DME positioning accuracy using the SFOL pulse, it would be best if the SFOL pulses could be transmitted from the DME transponders and interrogators currently being used in the field with minor software modifications. In a modern DME transmission unit, a baseline Gaussian pulse waveform stored in memory passes through multiple chains of pulse-shaping circuits and power amplifiers (PAs) before being transmitted, a process which may significantly distort the baseline pulse waveform. In most conventional DMEs, the waveform differences between the baseline Gaussian pulse and the actual transmitted pulse at the antenna are not properly corrected as long as the actual transmitted pulse still meets the ICAO DME pulse specifications. However, to fully utilize the multipath suppression capability of the SFOL pulse, the actual transmitted pulses must be as close as possible to the original SFOL pulse shape while satisfying the ICAO DME spectrum requirements. 

This letter proposes a digital predistortion (DPD) scheme for the transmission of an SFOL pulse from an existing Gaussian-pulse-based DME device. Similar to most DPD methods applied to PAs in mobile communications, our DPD approach is based on a memory polynomial representation, which models the nonlinearity and memory effect of a PA, and inverse learning algorithms. However, because the DME device contains a series of pulse-shaping circuits and PA modules, we noticed that the overall signal distortion characteristics of DME are different from those of PAs.

In this work, we first experimentally obtained the distortion patterns of the SFOL waveform using a commercial Gaussian-pulse-based DME transponder. Based on the obtained SFOL waveform distortion patterns, we propose a novel gain normalization method and introduce a bias term to the memory polynomial model for the predistortion of the SFOL waveform. The pulse shape difference between the original and transmitted SFOL waveforms was assessed using the root mean square (RMS) error, and it was confirmed that the transmitted waveform was very close to the original waveform while satisfying the ICAO DME pulse shape and spectrum requirements. In addition, the multipath resistance performance of the original and transmitted SFOL pulses was evaluated.

\section{SFOL DME Waveform Distortion Characteristics and Proposed Predistortion Method}
\begin{figure}
  \centering
  \includegraphics[width=0.9\linewidth]{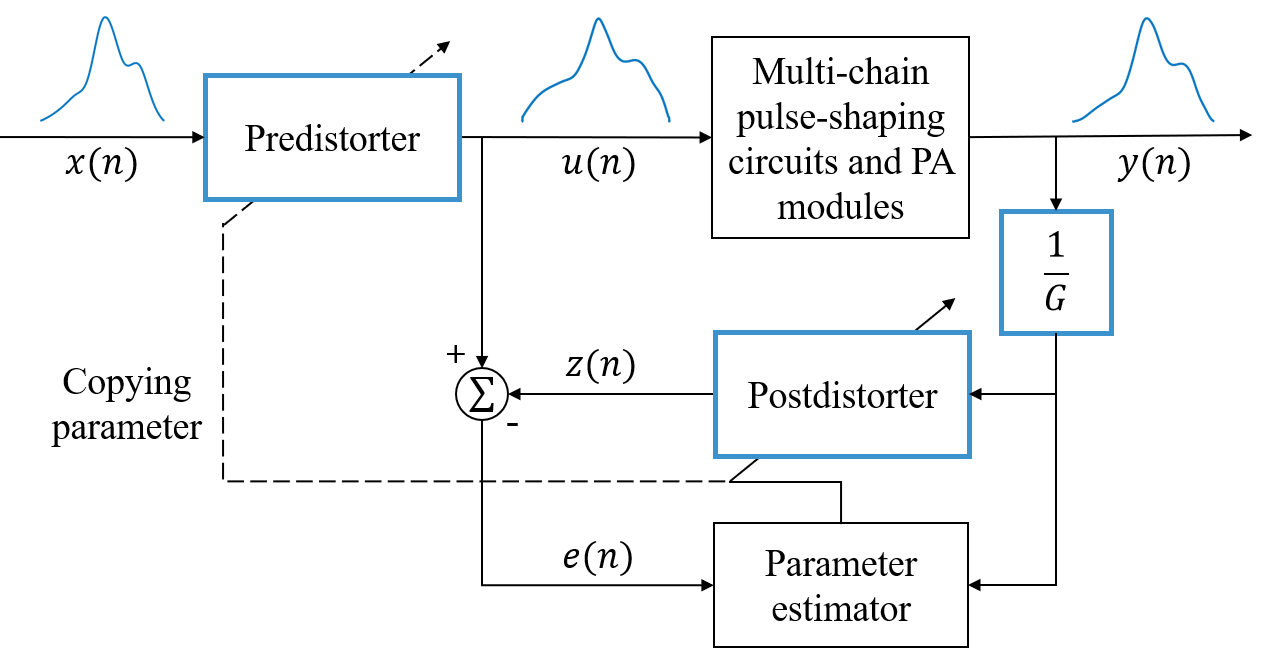}
  \caption{Block diagram of digital predistortion for DME. The gain normalization and predistorter/postdistorter blocks for PA predistortion were modified by the proposed methods to properly generate an SFOL DME pulse.}
  \label{fig:DPDscheme}
\end{figure}

One of the most widely used and cost-effective PA linearization techniques is digital predistortion, the overall procedure of which is depicted in Fig. \ref{fig:DPDscheme}. In Fig. \ref{fig:DPDscheme}, an RF input digital waveform $x(n)$ is predistorted before the PA. (The block with pulse-shaping circuits and PA modules for DME in the figure should be changed to a PA block when PA predistortion is considered.) When $u(n)$ passes through the PA, it experiences distortion owing to the nonlinearity of PA. The distorted waveform after PA is $y(n)$, and its normalized waveform with gain $G$ is fed into the postdistorter and parameter estimator to approximate the predistorter. This process is iterative, and the converged predistorter makes $y(n)$ closely follow $x(n)$, although the power level of $y(n)$ is larger than that of $x(n)$.

\subsection{Gain Normalization for DME}

\begin{figure}
  \centering
  \includegraphics[width=0.8\linewidth]{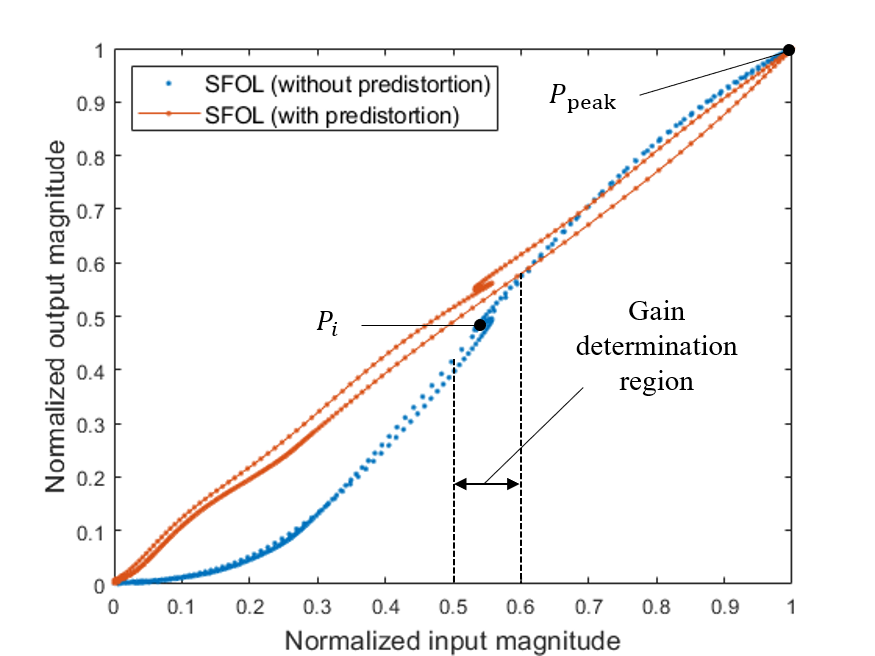}
  \caption{Normalized output vs. input magnitudes from experiments. The blue and red dots indicate the measurements from DME without and with predistortion, respectively. After applying the proposed predistortion method to DME, the input-output relationship was linearized.}
  \label{fig:Gain}
\end{figure}

As shown in Fig. \ref{fig:DPDscheme}, the PA output signal $y(n)$ is divided by the normalization gain $G$ before the postdistorter.
The normalization gain is typically selected as the maximum gain of the PA in its linear region or the gain at the maximum output power of the PA. Theoretically, both gain normalization methods exhibit the same performance \cite{Zhu06}. The normalization gain $G$ at the maximum output power, for example, is selected as $G_\mathrm{peak} = y_\mathrm{peak} / x_\mathrm{peak}$ at point $P_\mathrm{peak}$ in Fig. \ref{fig:Gain} for a PA, where $x_\mathrm{peak}$ and $y_\mathrm{peak}$ are the input and output magnitudes at point $P_\mathrm{peak}$, respectively. 
However, DME contains a series of pulse-shaping circuits and multiple PA modules, and its input-output magnitude relationship from our experiment was quite different from that of a PA, as shown in Fig. \ref{fig:Gain}. A typical PA shows linear behavior for low input magnitudes, but DME in our experiment showed nonlinear behavior, as in Fig. \ref{fig:Gain}, where the blue data points form a convex curve in the region of low input magnitudes. 
Thus, the typical gain normalization methods for PA would not provide optimal result when directly applied to digital predistortion for DME. 

\begin{figure}[t!]
  \centering
  \includegraphics[width=0.8\linewidth]{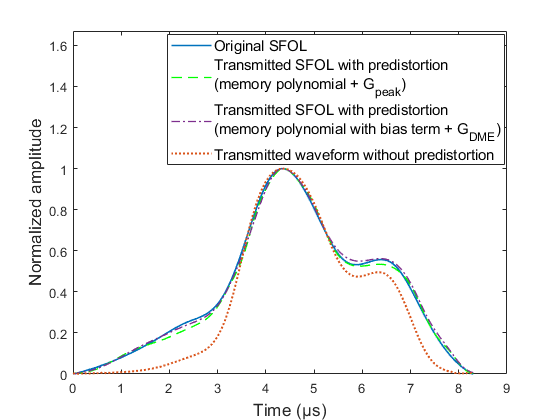}
  \caption{Comparison of pulse shapes of the original SFOL, transmitted SFOL after predistortion with $G_\mathrm{peak}$ and memory polynomial model, transmitted SFOL after predistortion with $G_\mathrm{DME}$ and memory polynomial with a bias term, and transmitted waveform without predistortion. The amplitude of each waveform is normalized by its peak value.}
  
  \label{fig:PlotWaves}
\end{figure}

Considering the nonlinear input-output relationship of DME in Fig. \ref{fig:Gain}, selecting the normalization gain $G$ is not straightforward. Because a single data point may not represent the average behavior of DME, we propose utilizing all data points $P_i$ in the gain determination region in Fig. \ref{fig:Gain} to calculate the normalization gain for DME predistortion. The proposed gain based on empirical data is expressed as $G_\mathrm{DME} = \sum{y_i} / \sum{x_i}$ over the gain determination region, where $x_i$ and $y_i$ are the input and output magnitudes at point $P_i$, respectively.

The gain determination region in Fig. \ref{fig:Gain} was selected as the region with high linearity from the operating curve of DME. First, a line was fitted with the data points in a candidate region using the least-squares method. The RMS error of the data points from the fitted line was then calculated. After obtaining the RMS errors of all candidate regions, the region with the minimum RMS error was selected as the gain determination region to be used for the $G_\mathrm{DME}$ calculation. A performance comparison between the two normalization gains (i.e., $G_\mathrm{peak}$ and $G_\mathrm{DME}$) during our field tests is presented in Section \ref{sec:Results}.

\subsection{Memory Polynomial with a Bias Term for DME} 

There are several methods to model the predistorter in Fig. \ref{fig:DPDscheme}, and the memory polynomial model is a widely applied model for PA predistortion because of its effectiveness and low computational complexity \cite{Morgan06, Tafuri12}. Once the memory polynomial model is applied, the input and output signals of the predistorter are related as follows: 
\begin{equation}
  u_\mathrm{MP}(n) = \sum_{k=0}^{K-1} \sum_{m=0}^{M-1} a_{km} x(n-m) {\lvert x(n-m) \rvert}^k
  \label{eqn:MP}
\end{equation}
where $K$ and $M$ are the nonlinearity order and memory depth, respectively, and $a_{km}$ are the model coefficients. This model considers not only nonlinearity but also the memory effects of PA.

However, we noticed additional important characteristics of DME compared to PA. When the SFOL waveform was input to DME without predistortion, the output waveform experienced significant distortion, as shown in the red curve in Fig. \ref{fig:PlotWaves}. One noticeable change in the output waveform (red) from the original SFOL waveform (blue) is the downward bias at the leading and trailing edges. 
When the time is less than 1 \si{\micro}s from the starting point of the SFOL waveform, for example, the normalized magnitude of the input signal (i.e., blue curve in Fig. \ref{fig:PlotWaves}) is less than 0.1. Then, the corresponding DME output signal has a normalized magnitude of less than 0.015 from the input-output relationship (blue) in Fig. \ref{fig:Gain}. As a result, the transmitted waveform without predistortion (the red curve in Fig. \ref{fig:PlotWaves}) has very small magnitudes when the time is less than 1 \si{\micro}s. Therefore, the observed downward bias of the transmitted waveform is due to the nonlinear behavior of DME in the region of low input magnitudes, which is not observed when using conventional PAs.

Because the conventional memory polynomial model in (\ref{eqn:MP}) does not have an explicit bias term, it may not properly consider the downward bias caused by DME. Therefore, we propose adding a bias term to the memory polynomial model as in (\ref{eqn:MPB}) for DME predistortion.
\begin{equation}
  u_\mathrm{MPB}(n) = \sum_{k=0}^{K-1} \sum_{m=0}^{M-1} a_{km} x(n-m) {\lvert x(n-m) \rvert}^k + b
  \label{eqn:MPB}
\end{equation}
where $b$ is the bias term used to model the DME bias effect. The parameters of this model are estimated using the feedback loop shown in Fig. \ref{fig:DPDscheme}.
The performance benefit from this model was experimentally demonstrated, as reported in the ensuing section.

\begin{figure}[t!]
  \centering
  \includegraphics[width=0.9\linewidth]{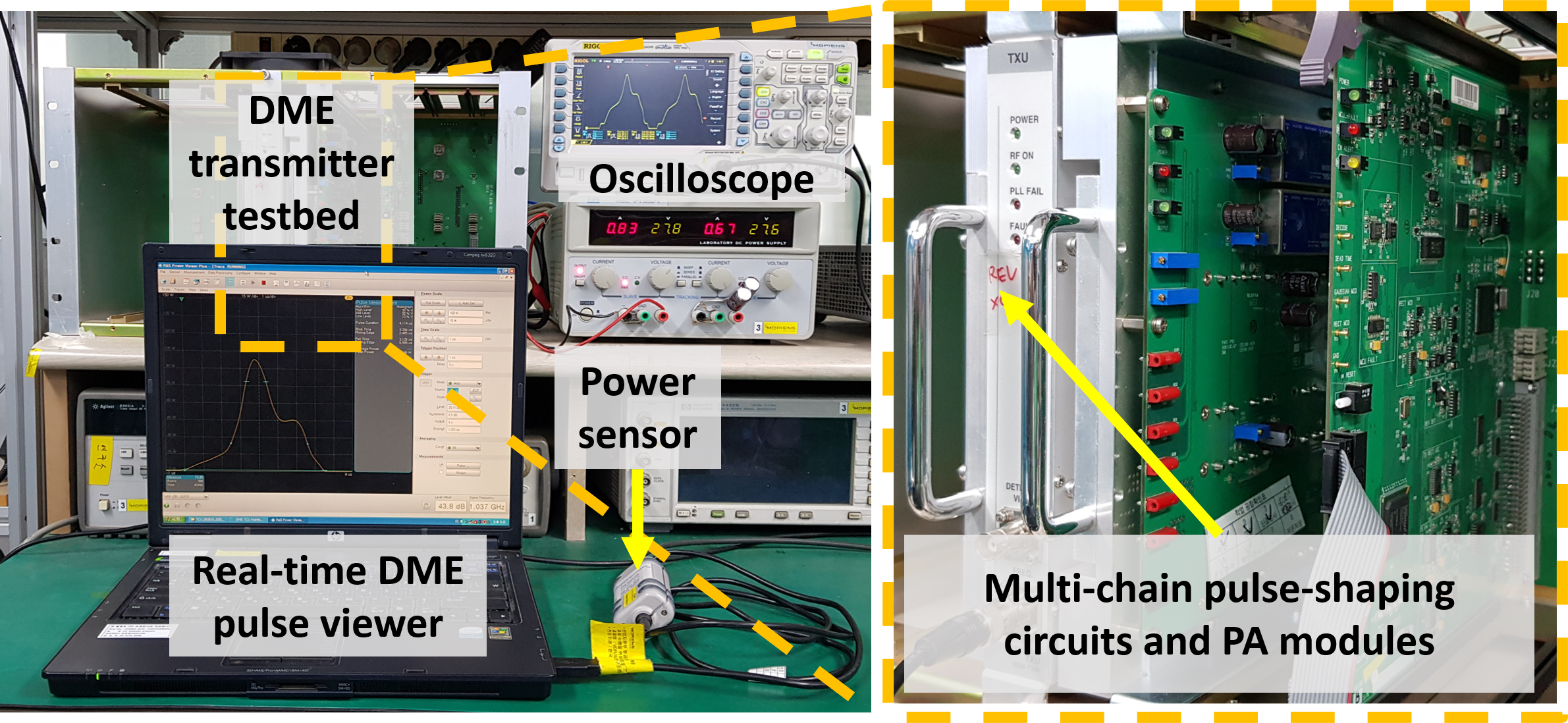}
  \caption{Experimental setup.}
  \label{fig:ExpSetup}
\end{figure}

\begin{figure}[t!]
  \centering
  \includegraphics[width=0.8\linewidth]{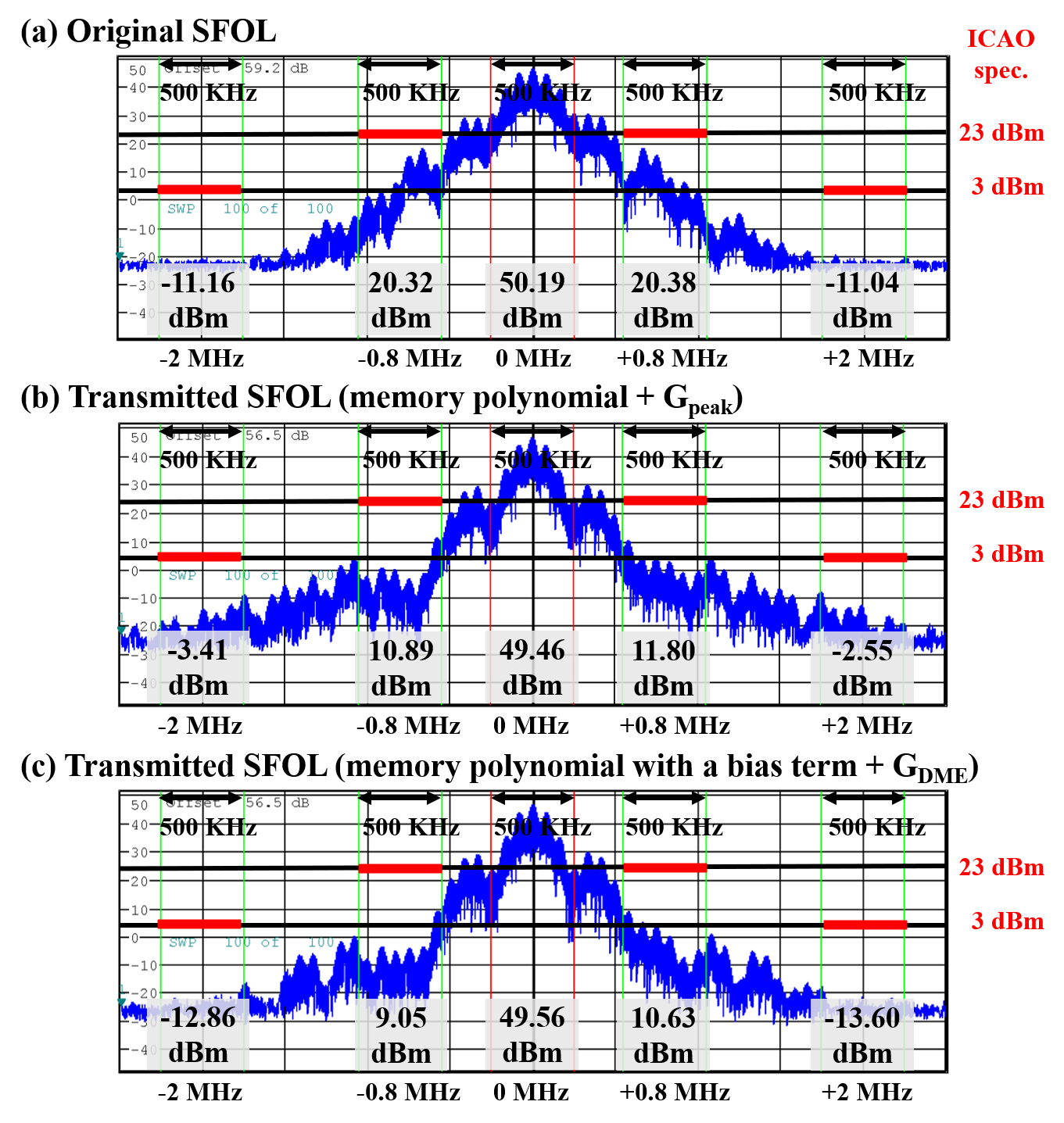}
  \caption{Captured images from the spectrum analyzer measuring power spectral density of: (a) transmitted waveform without predistortion, (b) transmitted SFOL after predistortion with $G_\mathrm{peak}$ and memory polynomial model, (c) transmitted SFOL after predistortion with $G_\mathrm{DME}$ and memory polynomial with a bias term.}
  \label{fig:PSD}
\end{figure}

\begin{table*}[t!]
\setlength{\tabcolsep}{0.32em}
\centering
\caption{RMS Error, Pulse Shape Parameters, and Channel Power}
\label{tab:RMSShapePower}
\begin{tabular}{@{\extracolsep{4pt}}cccccccccc} 
\hline\hline
\multirow{2}{*}{Normalization gain} & \multirow{2}{*}{Predistorter model} & \multirow{2}{*}{RMS error} & 
\multicolumn{3}{c}{Pulse shape parameters [\si{\micro}$s$]} & \multicolumn{4}{c}{$0.5$ MHz channel power [dBm]}  \\ 
\cline{4-6} \cline{7-10}
            &       &        
            & rise time & width & fall time
            & $-2$ MHz & $-0.8$ MHz & $+0.8$ MHz & $+2$ MHz            \\ 
\hline
$G_\mathrm{peak}$  & Memory polynomial        & $0.0134$
            & $2.89$      & $3.35$        & $3.04$
            & $-3.41$  & $10.89$    & $11.80$    & $-2.55$             \\
$G_\mathrm{peak}$  & Memory polynomial with a bias term       & $0.0072$          
            & $2.83$      & $3.37$        & $3.02$
            & $-7.73$  & $9.33$     & $11.32$    & $-6.96$             \\
$G_\mathrm{DME}$  & Memory polynomial        & $0.0114$          
            & $2.94$      & $3.42$        & $3.08$
            & $-8.39$  & $11.02$    & $11.28$    & $-8.28$             \\
$G_\mathrm{DME}$  & Memory polynomial with a bias term       & $0.0112$
            & $2.88$      & $3.46$        & $3.05$
            & $-12.86$ & $9.05$     & $10.63$   & $-13.60$             \\
\hline\hline
\end{tabular}
\end{table*}

\section{Field Test Results and Discussions}
\label{sec:Results}

Fig. \ref{fig:ExpSetup} shows our experimental setup for SFOL DME pulse shaping. The DME transmitter testbed taken from a commercial 100 W DME transponder from MOPIENS Inc. was set to transmit the SFOL pulse with a $1,037$ MHz channel frequency and $700$ Hz pulse repetition frequency. An oscilloscope was used to measure the waveform of $u(n)$, as shown in Fig. \ref{fig:DPDscheme}. A power sensor with a passive USB adapter acquired the transmitted pulse, $y(n)$, and sent it to a laptop computer. The laptop computer determined the normalization gain $G$ and estimated the predistorter/postdistorter parameters by the inverse-learning process.


To compare the performance of our proposed methods with that of the prior works, we tested four cases of DPD methods combining two gain normalization methods  (i.e., $G_\mathrm{peak}$ and $G_\mathrm{DME}$) and two predistorter models (i.e., memory polynomial and memory polynomial with a bias term) as listed in Table \ref{tab:RMSShapePower}. The nonlinearity order ($M$) and memory depth ($K$) were chosen as $2$ and $7$, respectively. The inverse learning process of the four cases converged at the eighth iteration at most.

Two performance measures are used for the transmitted SFOL pulses: pulse shape parameters and effective isotropic radiated power (EIRP) at selected frequencies. The original SFOL pulse shape has a rise time of 2.8 \si{\micro}s, a width of 3.4 \si{\micro}s, and a fall time of 3.0 \si{\micro}s. As listed in Table \ref{tab:RMSShapePower}, the pulse shape parameters of the transmitted SFOL pulses in all cases complied with the current ICAO DME specifications (i.e., rise time: less than 3 \si{\micro}s, width: 3.5 ($\pm0.5$) \si{\micro}s, fall time: less than 3.5 \si{\micro}s). Furthermore, the RMS errors of the transmitted waveforms compared with the original SFOL waveform were small for all four DPD cases. Fig. \ref{fig:PlotWaves} compares the transmitted pulse shapes of the two cases representing the prior works (i.e., $G_\mathrm{peak}$ and memory polynomial) and our proposed DPD method (i.e., $G_\mathrm{DME}$ and memory polynomial with a bias term). 

Regarding the EIRP, the current ICAO DME specifications allow the maximum EIRP contained in a $0.5$ MHz band centered at $\pm0.8$ MHz and $\pm2$ MHz away from a carrier frequency up to $23$ dBm and $3$ dBm, respectively. Fig. \ref{fig:PSD} shows the measured spectra of the three transmitted waveforms.
Because our testbed used a $100$ W DME device, the EIRP of the transmitted pulses ideally should be less than $13$ dBm and $-7$ dBm at $\pm0.8$ MHz and $\pm2$ MHz, respectively, to account for the $10$ dB EIRP increase with a $1,000$ W DME device. 
Among the four test cases, the case of $G_\mathrm{DME}$ and memory polynomial with a bias term has the largest EIRP margin with a small RMS error as shown in Table \ref{tab:RMSShapePower}; thus, it is our recommended DPD method for SFOL DME pulse shaping. Although the case using $G_\mathrm{peak}$ and the memory polynomial with a bias term showed the minimum pulse shape RMS error, it did not demonstrate an EIRP of less than $-7$ dBm at $+2$ MHz.

\begin{figure}[ht]
  \centering
  \includegraphics[width=0.9\linewidth]{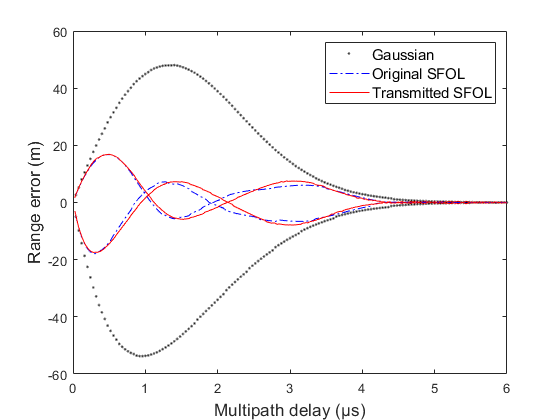}
  \caption{Comparison of multipath-induced range errors of the Gaussian, original SFOL, and transmitted SFOL pulses.}
  \label{fig:MultipathResult}
\end{figure}

The multipath mitigation performance of the transmitted SFOL pulse with the proposed DPD was compared with that of the Gaussian and original SFOL pulses through simulations as shown in Fig. \ref{fig:MultipathResult}. The peak ratio of the multipath and direct pulses was set to 0.3, and the delay of the multipath varied from 0 to 6 \si{\micro}s. The RMS of the multipath range errors of the Gaussian, original SFOL, and transmitted SFOL pulses were 18.5, 4.1, and 5.5 m, respectively. Although the multipath-induced range error of the transmitted SFOL pulse was slightly larger than that of the original SFOL pulse, it was significantly lower than that of the Gaussian pulse.

\section{Conclusion}
In this letter, we first presented DME characteristics and the corresponding SFOL waveform distortions that were experimentally obtained from commercial Gaussian-pulse-based DME hardware. To account for the nonlinear characteristics of DME, which are different from those of a typical PA, we proposed a novel DPD scheme that is applicable to a current Gaussian-pulse-based DME through software modifications alone. Our experimental tests confirmed that the proposed DPD method was able to transmit SFOL pulses with shapes nearly identical to the original SFOL pulse while satisfying the ICAO DME spectrum requirements. Furthermore, the multipath mitigation performance of the transmitted SFOL pulse demonstrated through simulation was satisfactory for a potential long-term APNT solution.   

\section*{Acknowledgment}

The authors gratefully acknowledge the support of Kwangwon Lee, Kyungsoo Woo, and Myungjoo Lee of MOPIENS Inc., Korea, for the preparation of the DME testbed and Suwoo Park of Hongik University for his help during tests.

\ifCLASSOPTIONcaptionsoff
  \newpage
\fi




\bibliographystyle{./IEEEtran}
\bibliography{./IEEEabrv,./mybibfile}

\end{document}